\documentclass[11pt,floatfix,aip]{revtex4-1}
\def \and{\& }

\def\by#1{#1,}
\def\and{and }
\def\yr#1{{(#1)}}
\def\paper#1{#1}
\def\paper#1{#1}
\def\jour#1{{\it #1}}

\def\vol#1{{\bf #1},}
\def\issue#1{}
\def\pages#1{\hbox{#1},}

\usepackage{amsmath}
\usepackage{amsfonts}
\usepackage{amssymb}
\usepackage{graphicx}
\usepackage{appendix}
\usepackage{placeins}

\begin{document}

\title{Surface permeability and surface flow tortuosity of particulate porous media.}
\author{Penpark Sirimark}
\author{Alex V. Lukyanov}
\email{corresponding author, a.lukyanov@reading.ac.uk}
\author{Tristan Pryer}
\affiliation{School of Mathematical and Physical Sciences, University of Reading, Reading RG6 6AX, UK}

\begin{abstract}
The dispersion process in particulate porous media at low saturation levels takes place over the surface elements of constituent particles and, as we have found previously by comparison with experiments, can be accurately described by super-fast non-linear diffusion partial differential equations. To enhance the predictive power of the mathematical model in practical applications, one requires the knowledge of the effective surface permeability of the particle-in-contact ensemble, which can be directly related with the macroscopic permeability of the particulate media.  We have shown previously that permeability of a single particulate element can be accurately determined through the solution of the Laplace-Beltrami Dirichlet boundary-value problem. Here, we demonstrate how that methodology can be applied to study permeability of a randomly packed ensemble of interconnected particles. Using surface finite element techniques we examine numerical solutions to the Laplace-Beltrami problem set in the multiply-connected domains of interconnected particles. We are able to rigorously estimate tortuosity effects of the surface flows in a particle ensemble setting.            
\end{abstract}

\maketitle
 
\section{Introduction}

Liquid transport in particulate porous media, such as sand, is customarily classified into fully saturated, funicular and pendular regimes of spreading~\cite{Bear-Book, Herminghaus-2005, Herminghaus-2008,  Herminghaus-2008-2}. The first two regimes of the liquid dispersion occur at relatively high saturation levels $s>s_c\approx 10\%$, where saturation $s$ is defined as the ratio of the liquid volume $V_L$ to the volume of available voids $V_E$ in a sample volume element $V$, $s=\frac{V_L}{V_E}$. At high saturation levels, above the critical value $s_c$, liquid transport takes place in the pore space either fully or partially filled by the liquid. 

\begin{figure}[ht!]
\begin{center}
\includegraphics[trim=-1.5cm 3.cm 1cm -0.5cm,width=\columnwidth]{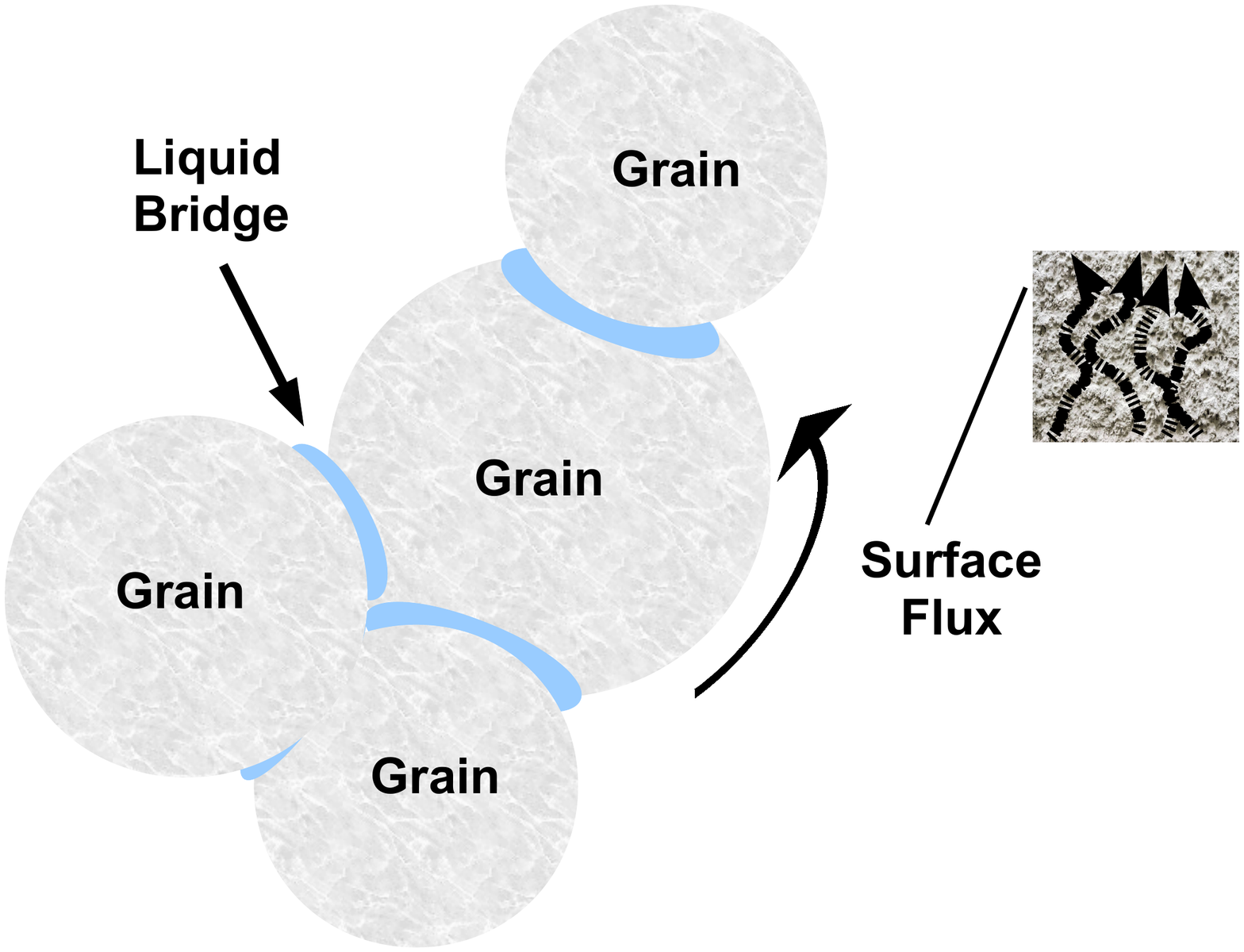}
\end{center}
\caption{Illustration of the liquid distribution in particulate porous media (grey) with pendular rings (blue) at low saturation levels.} 
\label{Fig1}
\end{figure}

Our prime concern here is the special case of liquid dispersion at low saturation levels. As the saturation level drops below the critical value, $s\le s_c$, that is to the value relevant to the pendular regime of spreading, the liquid volumes in the porous matrix become isolated~\cite{Herminghaus-2005, Herminghaus-2008, Herminghaus-2008-2}. As a result, at low saturation levels, the liquid is only contained in the pendular rings formed at the locations of the particle contacts and on the particle rough surfaces, and the liquid transport can only occur over the matrix surface elements, as is illustrated in Fig. \ref{Fig1}. 

Our main concern here is the wetting cycle, when the liquid spreads over a dry porous matrix or over a matrix with a very low background saturation level up to $s_r\approx 2\%$. These conditions are similar to those in the case studied previously experimentally and theoretically in~\cite{Lukyanov2019}. The main driving force of the dispersion process, as is often the case during the wetting cycle, is capillary pressure developed at the moving front in the process of wetting of dry porous matrix, while the liquid bridges play a role of variable liquid reservoirs of uniform surface curvature.  

The analysis of this regime of wetting, which is crucial for studies of biological processes and spreading of non-volatile liquids in arid natural environments and industrial installations, has shown that the liquid dispersion has many distinctive features and can be accurately described by the so-called superfast non-linear diffusion equation~\cite{Lukyanov2012, Lukyanov2019}.

Theoretically, the superfast non-linear diffusion equation belongs to a special class of mathematical models.  Unlike in the standard porous medium equation~\cite{Vazquez-Book}, in this special case, the non-linear coefficient of diffusion $D(s)$ demonstrates divergent behaviour as a function of saturation $s$, $D(s)\propto (s-s_0)^{-3/2}$, where $s_0$ is some minimal saturation level ($s_0\approx 0.5\%$), which could be only achieved in a state when the liquid bridges cease to exist completely~\cite{Lukyanov2012, Herminghaus-2008, Herminghaus-2008-2, Lukyanov2019}. Note, in that respect, that in the domain of spreading liquid bridges are supposed to never vanish, so that the condition  $s>s_0$ is always fulfilled in the model, and there is no actual singularity of the mathematical description~\cite{Lukyanov2012, Lukyanov2019}. 

Specifically, in the macroscopic approximation, that is after averaging over some volume element containing many particles of the porous medium, the diffusion process in the slow creeping flow conditions can be described by the following non-linear  diffusion equation 
\begin{equation}
\label{Superfast-1}
\frac{\partial s}{\partial t}= \nabla\cdot  \left\{ D(s) \nabla s \right\},  \quad t>0,
\end{equation}
where
$$
D(s)= \frac{D_0(s)}{(s-s_0)^{3/2}}, \quad s>s_0,
$$
for $D_0 > 0$.

The details of derivation of (\ref{Superfast-1}) can be found in~\cite{Lukyanov2012, Lukyanov2019}, here we note that, the resultant governing non-linear equation (\ref{Superfast-1}) directly follows from the conservation of mass principle
\begin{equation}
\label{mass-conservation}
\frac{\partial (\phi s)}{\partial t} + \nabla\cdot {\bf Q}=0,
\end{equation}
where $\phi$ is porosity defined as $\displaystyle \phi=\frac{V_E}{V}$, which is further assumed to be constant, and $\bf Q$ is the macroscopic flux density. The macroscopic flux density $\bf Q$ is defined in such a way that the total flux through the surface of a macroscopic sample volume element is given by the surface integral $\int {\bf Q\cdot n}\, dS$, where $\bf n$ is the normal vector to the surface of the sample volume element.

To obtain (\ref{Superfast-1}) from (\ref{mass-conservation}), one needs to apply the capillary pressure-saturation relationship~\cite{Halsey1998, Lukyanov2012, Lukyanov2019} dictated by the liquid bridges behaviour
\begin{equation}
\label{Pressure}
p = -p_0\frac{A_c}{(s-s_0)^{1/2}}
\end{equation}
and the local Darcy's law~\cite{Yost-1998, Tuller-2000} describing the surface flow in the rough layer of the particle elements
\begin{equation}
\label{Macroscopic-Darcy-2}
-\frac{\kappa_m}{\mu}  \nabla u ={\bf q}. 
\end{equation}

Here, $A_c=\sqrt{\frac{3}{4}\, \frac{1-\phi}{\phi} \frac{N_c}{\pi}}$, $N_c$ is the coordination number, that is the average number of bridges per a particle, $p_0=\frac{2\gamma}{R}\cos\phi_c$, $\gamma$ is the coefficient of the surface tension of the liquid, $\phi_c$ is the contact angle made by the free surface of the liquid bridge with the rough solid surface of the constituent particles, $R$ is an average radius of the porous medium particles, ${\bf q}$ and $u$ are the averaged local flux density and pressure in the rough surface layer, $\mu$ is liquid viscosity and $k_m$ is the local coefficient of permeability of the rough surface, which is proportional to the average amplitude of the
surface roughness $\delta_R$, that is the width of the surface layer conducting the liquid flux 
\begin{equation}
\label{kappam}
k_m\propto \delta_R^2.
\end{equation} 

One needs to emphasise here that two levels of averaging are involved in obtaining the final governing equation (\ref{Superfast-1}). While equations (\ref{Superfast-1}), (\ref{mass-conservation}) and (\ref{Pressure}) are 'truly' macroscopic, that is obtained by averaging using a volume element $V$ containing many grain particles, equation (\ref{Macroscopic-Darcy-2}) is only an average over some rough area of a single particle containing many surface irregularities, so that quantities ${\bf q}$ and $u$ are also only local averages over that sample surface area. 

Therefore, to transit from (\ref{Macroscopic-Darcy-2}) to the macroscopic description, the spatial averaging theorem formulated in~\cite{Whitaker-1969} should be applied. That is, using intrinsic liquid averaging $\langle ...\rangle ^l=V_l^{-1}\int_{V_l} d^3 x$, where $V_l$ is liquid volume within the sample volume $V$, one has $\langle u \rangle^l=p$ and $\langle {\bf q}\rangle^l \frac{S_e}{S} = {\bf Q}$. Here, $S$ is the surface area of the sample volume $V$ with the effective area of entrances and exits $S_e$. Note, the ratio $S_e/S$ is not just a geometric property, but also takes into account the connectivity of the porous elements. For example, the effective area of entrances and exits $S_e$ is only defined by the pathways open to the flow.

As a result of the two-level averaging
$$
D_0(s)=A_c \frac{K(s)}{\mu}\frac{p_0}{2\phi},
$$
where $K(s)=\kappa_m \frac{S_e}{S}$ is the coefficient of permeability defined by
$$
{\bf Q} = -\frac{K}{\mu}\nabla p.
$$

The global surface permeability of the particles $K$ as a function of saturation is one of the main elements of the model to accurately represent liquid dispersion at low saturation levels.  It is fully defined by the particle geometry and the geometry of the liquid bridge contact areas, Fig. \ref{Fig1} and Fig. \ref{Fig2}.

In particular, the disposition and the size of the liquid bridges on the particle surface, that is the size of the domains $\Omega_{1,2}$ and the angle $\alpha$, should play a leading role in defining the resistance to the surface flow. It is not difficult to discern that any variations of the contact area covered by the liquid bridges (pendular rings), that is areas $\Omega_{1,2}$ shown in Fig. \ref{Fig2}, or the value of the bridge volume, should affect the global permeability. 
 
Previously, we have shown that permeability of a single particle element can be determined by means of a solution to the equivalent Laplace-Beltrami boundary value problem formulated in the flow domain $\Omega_0$ with the boundaries $\partial \Gamma_{1,2}$ in Fig. \ref{Fig2}~\cite{Penpark2018}. We briefly formulate that problem and summarise the previous results in the next part. Here we note that, based on the analysis of the problem, we have been able to show that in a special azimuthally symmetric case of spherical particles, when the two areas covered by the liquid bridges, domains $\Omega_1$ and $\Omega_2$ in Fig. \ref{Fig2}, are oriented symmetrically to each other, that is at $\alpha=\pi$, the permeability $K$ is supposed to follow the scaling 
$$
K(s)\propto \frac{1}{|\ln(s-s_0)|}.
$$

We have studied several generalisations of the symmetric problem, such as arbitrary oriented domains, $\alpha\neq \pi$, on the surface of the spherical particle, and particles of arbitrary shapes emulating the shape of a real sand grain. While variations of the particle shape was found to produce a relatively modest effect on the particle surface permeability, the orientation of the boundaries, emulating tortuosity effects, was found to produce a stronger impact due to the substantial variation of the distance, on average, between the boundary contours $\partial \Gamma_{1,2}$. It became clear that while the previously obtained scaling was a good first step to estimate the surface permeability of particulate porous media, a more general case of an ensemble of interconnected particles should be analysed to enhance the model predictive power and at the same time to estimate rigorously the effects of tortuosity of the surface flow in the particle assembly. In this study, we will simulate a general case of an ensemble of many particles linked by liquid bridges. We will concentrate on the bunch of spherical particles, but of different radii and randomly arranged in configurations. We compare the random pack configuration results with some symmetric case to estimate the effects of tortuosity and formulate practical recipes to apply the super-fast diffusion model.

\section{Microscopic model of the surface permeability of the elements}

Microscopically, the liquid creeping flow through the surface roughness of each particle can be described by a local Darcy-like relationship (\ref{Macroscopic-Darcy-2}) between the surface flux density ${\bf q}$ and averaged (over some area containing many surface irregularities) pressure in the grooves $u$~\cite{Yost-1998, Tuller-2000}. Assuming incompressibility of the liquid and that the liquid layer thickness is constant $\delta_R=const$, one has 
\begin{equation}
\label{Incomp}
\nabla\cdot {\bf q}=0.
\end{equation}

Equation (\ref{Macroscopic-Darcy-2}) taking into account (\ref{Incomp}) can then be transformed into the Laplace-Beltrami equation defined on the surface $\Gamma$ of the particle
\begin{equation}
\label{Laplace-Beltrami-ini}
\Delta_{\Gamma} u =0.
\end{equation} 
Here, $\Delta_{\Gamma}$ designates the Laplace-Beltrami operator, which is defined on the surface element $\Gamma$ through the surface gradient $\nabla_{\Gamma}$ tangential to the surface. Formally, let ${\bf{n}}_{\Gamma}$ denote the unit normal to the surface $\Gamma$, Fig. \ref{Fig2}. Then, one can define the surface gradient of a smooth function $u$ as $\nabla_{\Gamma} u:=\nabla u - (\nabla u \cdot {\bf{n}}_{\Gamma}) {\bf{n}}_{\Gamma}$ and then the Laplace-Beltrami operator is defined as $\Delta_{\Gamma} u = \nabla_{\Gamma}\cdot \nabla_{\Gamma} u$.

The second assumption $\delta_R=const$ implies that the surface layer is fully saturated, that is its content is not changing on the particle surface. The approximation of the fully saturated rough surface layer is well fulfilled, if the characteristic pressure amplitude $|u|$ is less than the capillary pressure amplitude defined on the length scale of the surface roughness $\delta_R$, which is of the order of $\delta_R\sim 1\,\mu\mbox{m}$ in typical sands \cite{Alshibli2004}, as is demonstrated in~\cite{Yost-1998}. That is, $|u| < u_c=\frac{\gamma}{\delta_R}$, and, for example for water ($\gamma=72\, \mbox{mN}/\mbox{m}$) at $\delta_R =1\,\mu\mbox{m}$, this results in $|u| < 7.2\times 10^4\, \mbox{Pa}$. 

Alternatively, if the surface layer somehow is not fully saturated, parameter $\delta_R$ should be interpreted as the characteristic width of the liquid layer within the rough surface layer and one needs to presume that variations of the pressure $|\delta u|$ are negligible $|\delta u|\ll u_c$. This is usually the case in slow, creeping flow conditions in porous media, and in fact, it is a criterion for the use of macroscopic approximation to such flows~\cite{Bear-Book}. As is shown in~\cite{Lukyanov2019}, strong negative capillary pressure on the level of $u_c$ are only expected at the moving front, so that the approximation is well fulfilled in the macroscopic flow domain. Note also that, it is always assumed throughout this study that 
$$
\delta_R\ll R,
$$ that is the amplitude of the surface roughness (or the width of the liquid layer) is always much smaller than the particle size.

\subsection{Permeability of a single  particle element}

\begin{figure}[ht!]
\begin{center}
\includegraphics[trim=-1.5cm 1.cm 1cm -0.5cm,width=0.6\columnwidth]{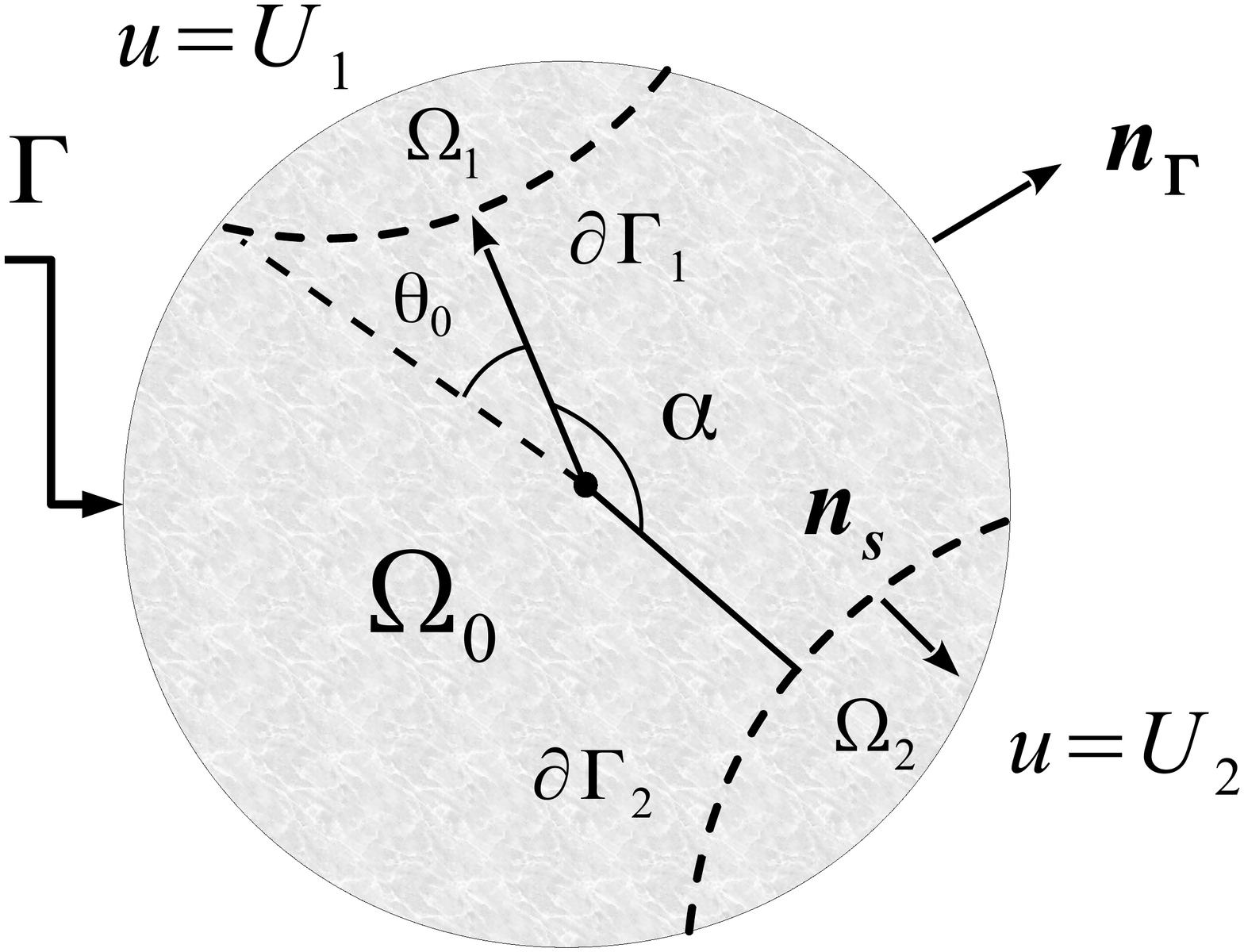}
\end{center}
\caption{Illustration of the flow and solution domains on the surface $\Gamma$ of a spherical particle, and their geometric arrangements. In the picture, $\Omega_0$ is the domain of the surface flow and the surface area covered by the liquid bridges corresponds to the domains $\Omega_1$ and $\Omega_2$.} 
\label{Fig2}
\end{figure}

Consider, as the simplest example, a spherical particle of radius $R$ with a closed surface $\Gamma$, which is split into three sub-domains $\Omega_0$, $\Omega_1$ and $\Omega_2$ with the surface boundaries between them $\partial \Gamma_1$ and $\partial \Gamma_2$, as is shown in Fig. \ref{Fig2}. The location of the sub-domains $\Omega_1$ and $\Omega_2$ to each other on the surface is fixed by the tilt angle $\alpha$. The sub-domains $\Omega_1$ and $\Omega_2$ correspond to the contact area covered by the liquid in the bridges, while the surface flow, described by (\ref{Macroscopic-Darcy-2}), takes place in $\Omega_0$. Our prime concern is permeability of the surface elements, so that we only consider steady state problems. 

The distribution of liquid pressure $u$, as it follows from (\ref{Laplace-Beltrami-ini}), should satisfy the Laplace-Beltrami equation now defined on the surface of the sub-domain $\Omega_0$
\begin{equation}
\label{Laplace-Beltrami}
\Delta_{\Omega_0} u =0.
\end{equation}   

Note that, in fact, the condition of the fully saturated surface layer is not essential in calculation of the flows over one particle element of the porous media. It is sufficient to presume that the variation of the capillary pressure on the length scale of the particle $|\delta u|$ is negligible, that is $|\delta u | \ll u_c$.  In the case when the surface layer is not fully saturated, parameter $\delta_R$ should be interpreted as the effective thickness of the layer filled by the liquid.

At the same time, liquid pressure variation in the bridges is negligible in slow, creeping flows in comparison to that in $\Omega_0$. So that, one can assume that
\begin{equation}
\label{Laplace-Beltrami-BCS}
\left. u \right|_{\partial \Gamma_1}=U_1=const, \quad \left. u\right|_{\partial \Gamma_2}=U_2=const,
\end{equation}
which are the boundary conditions to the Laplace-Beltrami Dirichlet boundary value problem. The Dirichlet boundary value problem (\ref{Laplace-Beltrami})-(\ref{Laplace-Beltrami-BCS}) has at least a unique weak solution, if the domain $\Omega_0$ and the boundaries $\partial \Gamma_{1,2}$ are smooth~\cite{Dziuk1988, AMS2005, Dziuk2013}, which, if it is found, allows to calculate the total flux through the particle element
\begin{equation}
\label{TotalInt}
Q_T=\delta_R\frac{\kappa_m}{\mu}\int_{\partial \Gamma_1} \frac{\partial u}{\partial n_s}\, dl =  -\delta_R\frac{\kappa_m}{\mu} \int_{\partial \Gamma_2} \frac{\partial u}{\partial n_s} \, dl,
\end{equation}
where $\bf n_s$ is the normal vector to the domain boundaries $\partial \Gamma_{1,2}$ on the surface, $\delta_R$ is the average amplitude of the surface roughness, that is the width of the surface layer conducting the liquid flux and the line integral is taken along a closed curve in $\Omega_0$, for example the boundary $\partial \Gamma_1$. 

If the total flux $Q_T$ is determined, one can define the global permeability coefficient of a single particle $K_1$. This can be done, if we assume that the particle has a characteristic size $D$ and so that it can be enclosed in a volume element $V=D^3$ with the characteristic side surface area $D^2$. Then, the effective flux density $Q$ can be represented in terms of $K_1$ (and the total flux $Q_T$)
\begin{equation}
\label{TotalFDef}
Q=\frac{Q_T}{D^2}=-\frac{K_1}{\mu}\frac{U_2-U_1}{D},
\end{equation}
if the flow is driven by the constant pressure difference $U_2-U_1$ applied to the sides of the volume element.

\subsection{Surface permeability of a sphere in the case of azimuthally symmetric domain boundaries}

Consider now a spherical particle in an azimuthally symmetric case, when the domain boundaries $\partial \Gamma_1$ and $\partial \Gamma_2$ are oriented at the reflex angle $\alpha=\pi$ and have a circular shape. We use a spherical coordinate system with its origin at the particle centre and the polar angle $\theta$ counted from the axis of symmetry passing through the centre of the circular contour $\partial \Gamma_1$. In this case, the Dirichlet boundary value problem (\ref{Laplace-Beltrami})-(\ref{Laplace-Beltrami-BCS}) admits an analytical solution, so that particle permeability can be determined explicitly. Indeed,
problem (\ref{Laplace-Beltrami})-(\ref{Laplace-Beltrami-BCS}), if we assume that the liquid pressure distribution $u$ is a function of $\theta$ only and independent of the azimuthal angle, is equivalent to
\begin{equation}
\label{Laplace-Beltrami-symmetric}
\frac{1}{\sin\theta}\frac{\partial }{\partial \theta}\left( \sin\theta \frac{\partial u}{\partial \theta}\right)=0, \quad \theta_0 < \theta < \pi-\theta_1,
\end{equation}
with the boundary conditions
\begin{equation}
\label{BCLB-symmetric}
\left. u\right|_{\theta=\theta_0}=U_1, \quad \left. u \right|_{\theta=\pi-\theta_1}=U_2.
\end{equation}

The analytic solution to problem (\ref{Laplace-Beltrami-symmetric})-(\ref{BCLB-symmetric}) after applying the boundary conditions can be represented in the following form 
\begin{equation}
\label{Laplace-Beltrami-Analytic}
u=\Psi_0 (U_2-U_1) \ln \left\{\frac{\sin\theta}{\sin\theta_0} \frac{1+\cos\theta_0}{1+\cos\theta}  \right\} +U_1,
\end{equation}
where
$$
\Psi_0=\frac{1}{\ln \left\{\frac{\sin\theta_1}{\sin\theta_0} \frac{1+\cos\theta_0}{1-\cos\theta_1}  \right\}}.
$$

One can now calculate the total flux and the permeability, using its definition (\ref{TotalFDef}), 
$$
Q_T=-\frac{K_1}{\mu} D (U_2-U_1)=-2\pi \sin\theta_0 \delta_R \frac{k_m}{\mu}\left. \frac{\partial u}{\partial \theta}\right|_{\theta=\theta_0}
$$
\begin{equation}
\label{QT1}
=-(U_2-U_1) 2\pi \delta_R \Psi_0 \frac{k_m}{\mu}. 
\end{equation}
So that, taking $D=2R$, 
\begin{equation}
\label{PSphere}
K_1=\pi \Psi_0 \frac{\delta_R}{R} k_m.
\end{equation}

Parametrically, the coefficient of permeability (\ref{PSphere}) is inversely proportional to the particle radius $R$, so that larger particles create stronger resistance to the flow. Noticeably, the coefficient demonstrates strong dependence on the surface layer thickness $\delta_R$, that is $K_1\propto \delta_R^3$ since it is anticipated that $k_m\propto \delta_R^2$, so that evaluation of this parameter in applications is crucial for the accurate estimates of the liquid dispersion rates.

One can see, if we take $\theta_1=\theta_0$, in fact assuming small variations of the bridge size and the pressure over one particle diameter, and $\theta_0\ll 1$, in fact considering small values of saturation, $s\ll 1$, that the permeability coefficient $K_1$ tends to zero, that is
\begin{equation}
\label{K1D}
K_1 = \frac{\delta_R}{2R} \frac{\pi k_m}{|\ln\theta_0 |} + o\left(\frac{1}{|\ln\theta_0|}\right).
\end{equation}

How does the result affect the super-fast diffusion model (\ref{Superfast-1}), and basically how can it be incorporated into the main diffusion equation? If we approximate the permeability coefficient $K$ by $K_1$ obtained in the azimuthally symmetric case at $\theta_1=\theta_0$, (\ref{K1D}), and, using an approximate relationship between the radius of curvature $R\sin\theta_0$ of the boundary contour $\partial \Gamma_1$ and the pendular ring volume~\cite{Herminghaus-2005}, one can show that
$$
\sin^2\theta_0 \approx \theta_0^2 = \sqrt{s-s_0}. 
$$
Therefore, finally
\begin{equation}
\label{GPSphere}
K(s)\approx 2 \frac{\delta_R}{R} \frac{\pi k_m}{|\ln(s-s_0)|}.
\end{equation}
As it follows from (\ref{GPSphere}), the distinctive particle shape results in logarithmic correction to the main non-linear superfast-diffusion coefficient $D(s)=\frac{D_0(s)}{(s-s_0)^{3/2}}$, such that 
$$D(s)\propto \frac{1}{|\ln(s-s_0)|(s-s_0)^{3/2}}.$$ Apparently, the correction will mitigate to some extent the divergent nature of the dispersion at the very small saturation levels $s\approx s_0$, smoothing out the characteristic dispersion curves. 

\subsection{Surface permeability of a chain of spheres in the case of azimuthally symmetric domain boundaries}

\begin{figure}[ht!]
\begin{center}
\includegraphics[trim=-1.5cm 0.2cm 1cm 0.5cm, width=0.5\columnwidth]{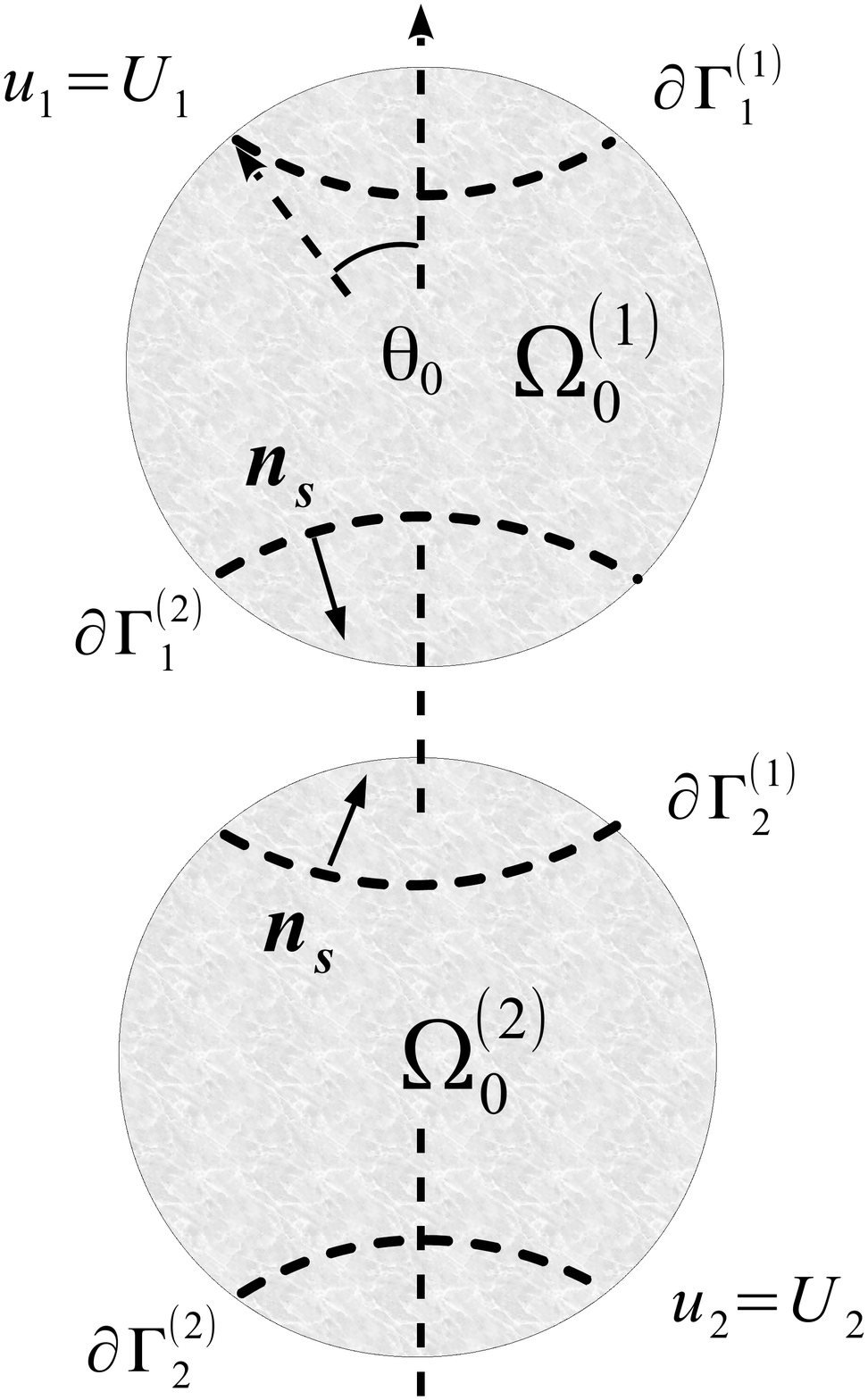}
\end{center}
\caption{Illustration of the solution domains in a system of two coupled spherical particles and their geometric arrangements.} 
\label{Fig3}
\end{figure}

Consider now how the problem can be formulated in the case of several particles arranged in a single chain, as is illustrated in Fig. \ref{Fig3} in the case of two coupled by the bridge particles. To create the flow in the system of two coupled particles, one can set pressure difference between $\partial \Gamma_1^{(1)}$ and $\partial \Gamma_2^{(2)}$. Mathematically, this is equivalent of setting Dirichlet boundary conditions on $\partial \Gamma_1^{(1)}$ and $\partial \Gamma_2^{(2)}$ as in the previous case of a single particle. The boundaries $\partial \Gamma_1^{(2)}$ and $\partial \Gamma_2^{(1)}$ are 'internal', that is common to the bridge linking the flow between the two particles. Apparently, the pressure is supposed to be the same on the two contours 
\begin{equation}
\label{Cont1}
u_1\left. \right|_{\partial \Gamma_1^{(2)}}=u_2\left. \right|_{\partial \Gamma_2^{(1)}}=const
\end{equation}
and due to conservation of mass in steady state conditions in the absence of sinks and sources of the liquid one has 
\begin{equation}
\label{Cont2}
\oint_{\partial\Gamma_1^{(2)}} \nabla u_1 \cdot  {\bf n}_{s_1}|_{\partial\Gamma_{1}^{(2)}}\, dl =-\oint_{\partial\Gamma_2^{(1)} }\nabla u_2 \cdot  {\bf n}_{s_2}|_{\partial\Gamma_{2}^{(1)}}\, dl
\end{equation}
where ${\bf n}_{s_1}$ and ${\bf n}_{s_2}$ are the outward tangential normal vectors to the boundary contours $\partial\Gamma_{1,2}^{(2,1)}$, and $u_1$ and $u_2$ designate distribution of pressure on each particle respectively.

As a result, the problem to define the flow and the permeability of the system corresponds to a system of two Laplace-Beltrami equations 
\begin{equation}
\label{LBS2p1}
\frac{1}{\sin \theta} \frac{\partial}{\partial \theta} \left ( \sin \theta \frac{\partial u_1}{\partial \theta} \right ) = 0 \ \ ,  \ \ \theta_0 \leq  \theta \leq  \pi - \theta_0
\end{equation}
and
\begin{equation}\label{LBS2p2}
\frac{1}{\sin \theta} \frac{\partial}{\partial \theta} \left ( \sin \theta \frac{\partial u_2}{\partial \theta} \right ) = 0 \ \ , \ \ \theta_0 \leq  \theta \leq  \pi - \theta_0,
\end{equation}
but with a slightly different set of the boundary conditions
\begin{equation}\label{LBS2p2B1}
u_1 \left.\right |_{\theta=\theta_0} = U_1
\end{equation}  

\begin{equation}\label{LBS2p2B2}
u_2 \left.\right |_{\theta=\pi-\theta_0}  =  U_2,
\end{equation}

\begin{equation}\label{LBS2p2B3}
u_1 \left.\right |_{\theta=\pi-\theta_0} =  u_2 \big |_{\theta=\theta_0}
\end{equation}
and
\begin{equation}\label{LBS2p2B4}
\left. \left(\sin \theta \dfrac{\partial u_1}{\partial \theta} \right) \right|_{\theta=\pi-\theta_0} =  \hspace{0.5cm} \left. \left( \sin \theta \dfrac{\partial u_2}{\partial \theta}\right) \right |_{\theta=\theta_0},
\end{equation}
where $\theta_0$, as before, defines the size of the bridge footprint on the particle surface in the spherical coordinate system with the axis of symmetry passing through the centre of the bridge area, Fig. \ref{Fig3}. Since we assumed, due to relatively small variations of pressure over a few grain particles, that all bridges are roughly identical, we have only one parameter $\theta_0$ to describe the bridge size.  

Apparently, equations (\ref{LBS2p1}) and (\ref{LBS2p2}) can be integrated twice, similar to the previous problem of a single particle (\ref{Laplace-Beltrami-symmetric}), to obtain
\begin{equation} \label{GS2P1}
 u_{1} = C_0 \ln \frac{\sin \theta}{1+\cos \theta} + C_1,  
\end{equation}

\begin{equation}  \label{GS2P2}
u_{2} = B_0 \ln \frac{\sin \theta}{1+\cos \theta} + B_1,   
\end{equation}
where $C_{1,2}$ and $B_{1,2}$ are free constant parameters to be found from the boundary conditions.

It is not difficult to see from (\ref{LBS2p2B4}), that one has $C_0=B_0$ implying continuity of the contact flux. Applying the remaining boundary conditions (\ref{LBS2p2B1})-(\ref{LBS2p2B3}), from (\ref{GS2P1}) and (\ref{GS2P2}) 
\begin{equation} \label{S2P1}
 u_{1} = \Psi_0^{(2)}  (U_2-U_1)  \ln\left( \frac{\sin \theta}{1+\cos \theta} \frac{1+\cos \theta_0}{\sin \theta_0} \right) + U_1,  
\end{equation}
\begin{equation}  \label{S2P2}
u_{2} = \Psi_0^{(2)} (U_2-U_1) \ln \left( \frac{\sin \theta}{1+\cos \theta} \frac{1 - \cos \theta_0}{\sin \theta_0} \right)  + U_2,   
\end{equation}
where
$$
\Psi_0^{(2)}=\frac{1}{2\ln\left(\frac{1+\cos\theta_0}{1-\cos\theta_0}\right)}.
$$

One can now calculate total flux and define permeability of the coupled spherical particles $K_2$
$$
Q_T=-\frac{K_2}{2\mu} D (U_2-U_1)=-2\pi \sin\theta_0 \delta_R \frac{k_m}{\mu}\left. \frac{\partial u_1}{\partial \theta}\right|_{\theta=\theta_0}
$$
\begin{equation}
\label{QT2}
=-(U_2-U_1) 2\pi \delta_R \Psi_0^{(2)} \frac{k_m}{\mu}, 
\end{equation}
where $D$ is the characteristic length scale of the cross-section in the problem, $\kappa_m$ is local permeability of the surface layer, $\delta_R$ is the layer width and $\mu$ is liquid viscosity.

So that, taking simply $D=2R$, 
\begin{equation}
\label{CPSphere}
K_2=2\pi \Psi_0^{(2)} \frac{\delta_R}{R} k_m.
\end{equation}

One can see that the permeability of a system of two coupled particles $K_2$ is identical to that of a single particle (\ref{PSphere}), basically from (\ref{PSphere}) and (\ref{CPSphere})
$$
\frac{K_2}{K_1}=\frac{2\Psi_0^{(2)}}{\Psi_0}=1.
$$

It is not difficult to discern by deduction that in a general case of $N$ coupled particles in a chain 
\begin{equation}
\label{CCPSphere}
K_N=\pi N \Psi_0^{(N)} \frac{\delta_R}{R} k_m=K_1
\end{equation}
where
$$
\Psi_0^{(N)}=\frac{1}{N\ln\left(\frac{1+\cos\theta_0}{1-\cos\theta_0}\right)}.
$$
Note, experimentally, the setup of many beads coupled by liquid bridges is often used in microfluidics to create flexible water channels~\cite{Nanoscale2016}. If the radius of curvature of the particle chain is much larger than the particle size, the transport through such a microfluidic system should be defined by the permeability of a single particle, relationship (\ref{PSphere}), if the particle shape can be approximated by a sphere. 

\begin{figure}[ht!]
\begin{center}
\includegraphics[trim=0.2cm 1.2cm 0.2cm 0.5cm, width=0.5\columnwidth]{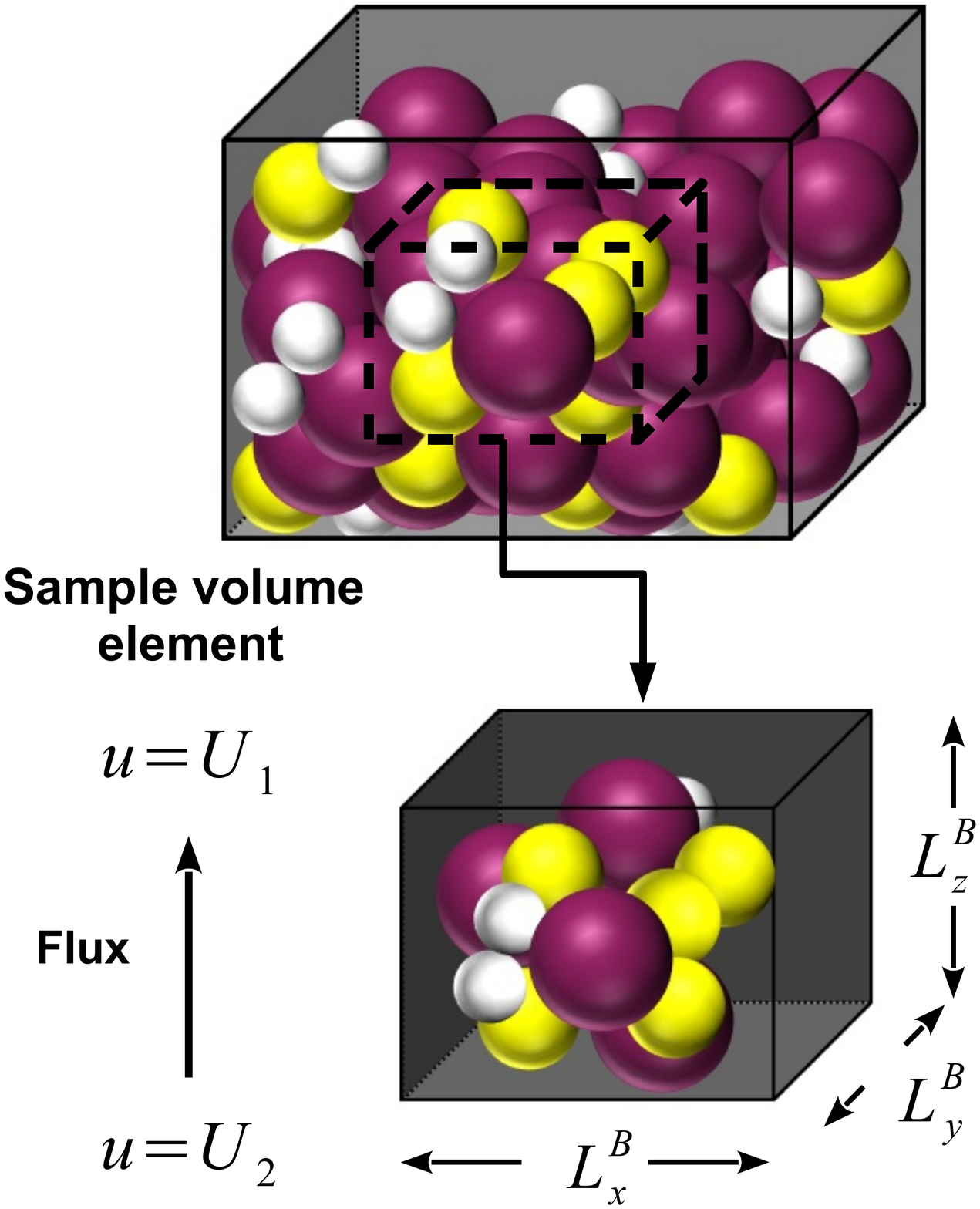}
\end{center}
\caption{Schematic illustration of the particle ensemble and the sample volume element setup.} 
\label{Fig4}
\end{figure}

One can conclude in this part, that if the porous media configuration is made of parallel chains of particles oriented symmetrically to each other, and the flow is generated along the chains, the surface permeability given by (\ref{PSphere}) is the exact result. 

\section{Surface permeability of a randomly packed particle ensemble}

In real systems, the particles are interconnected randomly, so that the effects of tortuosity should substantially affect the permeability of the system~\cite{Tortuosity1937, Tortuosity1961, Bear-Book, Tortuosity-Review2013}. To analyse those effects, we consider an ensemble of spherical particles randomly packed, as is shown in Fig. \ref{Fig4}. The randomly packed configuration of approximately $3000-7000$ particles has been generated by means of a molecular dynamics technique by applying a constant force to every particle placed in a box with reflecting boundaries (in the perpendicular direction to the box side), and interacting via the Lennard-Jones potential with different characteristic length scales $R$ distributed normally, that is with the probability of the particle radius $W(R)\propto \exp\left(-\frac{(R-R_0)^2}{\Delta R^2}\right)$ at $\Delta R/R_0 = 0.3$. In this study, there were particles with three different characteristic dimensions $R_1=1.3\, R_0$, $R_2=R_0$ and $R_3=0.7\, R_0$. The resultant porosity in the configurations was about $48\%$.

To obtain the configuration, the particle temperature controlled by the thermostat has been gradually reduced to bring the system to a minimum energy, frozen state. A representative sample volume element with dimensions $L_x^B, L_y^B, L_z^B$ then was cut off the system, as is illustrated in Fig. \ref{Fig4}, containing $N_S=13-17$ particles, see Table \ref{Table1} for details. We have generated several statistically independent sample configurations, and, as in the previous examples, set constant pressure difference $U_2-U_1$ at the boundaries of the sample elements, Figs. \ref{Fig4} and \ref{Fig6}. 

The Laplace-Beltrami method then has been applied after establishing the position of the liquid bridges coupling the particles in the sample. Two particles (of radii $R_1$ and $R_2$) are assumed to be coupled by a liquid bridge if the distance between their centres $r$ was only slightly larger than the sum of their radii
$$
R_1+R_2 \le r < R_1 + R_2 +0.05\max(R_1,R_2).
$$

\begin{figure}[ht!]
\begin{center}
\includegraphics[trim=0.2cm 1.cm 0.2cm 0.5cm, width=0.5\columnwidth]{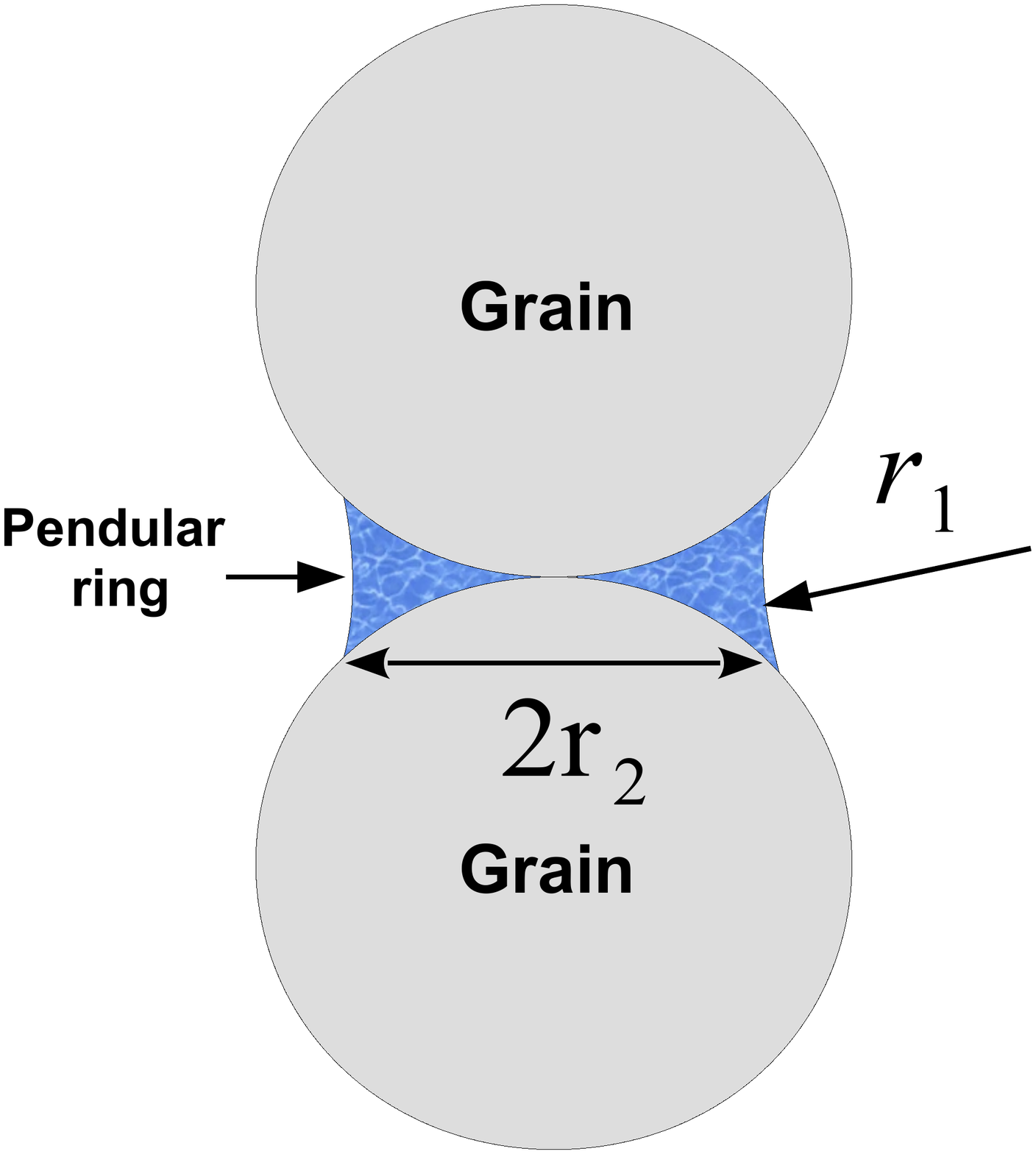}
\end{center}
\caption{Illustration of the pendular ring characteristic geometry.} 
\label{FigB}
\end{figure}

The size of a single liquid bridge footprint $H_B$ on the particle surface can be characterised, as before, by the polar angle $\theta_0$ in the polar coordinate system with the symmetry axis passing through the centre of the circular contour, the boundary of the area covered by the bridge, as is shown in the symmetric case in Fig. \ref{Fig3}. That is, $H_B=2R_k\sin\theta_0^{(k)}$. Due to the specific geometric properties of the pendular rings (constant mean curvature surface), we assume that even in the case of a distribution of particles with different radii $R_k$, the size of the bridge area in the sample is approximately the same in the low saturation limit $s\ll 1$ ($\theta_0^{(k)}\ll 1$)~\cite{Herminghaus-2005, Halsey1998, Willett2000}. 

Indeed, when $s\ll 1$, the pressure in the pendular ring $p$ is defined by the smallest radius of curvature $r_1$, Fig. \ref{FigB}, $p\approx - \gamma\cos\phi_c/r_1$, which is related with the second radius 
\begin{equation}
\label{BridgeS}
r_1\approx r_2^2/2R_k,
\end{equation}
so that when $s\ll 1$, one has $r_2\gg r_1$. Obviously, $r_2$ defines the size of the area covered by the bridge, $H_B=2r_2=2R_k\sin\theta_0^{(k)}$. 

If we have two particles of different radii, say $R_1$ and $R_2$, in contact, the size of the bridge area will be approximately the same $r_2^{(1)}\approx r_2^{(2)}$ at low saturation levels, $s\ll 1$, with the difference being proportional to $r_1$, that is
\begin{equation}
\label{Bridgetwo}
\frac{ r_2^{(1)} - r_2^{(2)} }{ \max(R_1,R_2) }=O\left( \frac{r_1}{\max(R_1,R_2)} \right).
\end{equation}

Apparently, in a general case, no analytic solution is expected to the Laplace-Beltrami problem and a well established surface finite element technique~\cite{Dziuk1988, AMS2005, Dziuk2013} is applied after the tessellation of the domains, as is shown in Fig. \ref{Fig5} for one particulate element with two boundary contours. The numerical method has been validated against analytical solutions previously demonstrating prescribed order of accuracy and numerical convergence, see details in~\cite{Penpark2018}.

The number of particles in the sample volume element was negotiated between computational efficiency of the surface finite element method (so that, practically, any mesh resolution can be used to deal with any details on the boundary contours $\partial \Gamma_{k}^{(l)}$ and on the particle surfaces) and fluctuations of the averaged quantities obtained using the sample element, which are proportional to $N_S^{-1/2}\approx 25\%$. Moderate increase of the number of particles in the sample may significantly increase computational time to obtain highly resolved numerical solutions, while at the same time would not substantially reduce the effect of particle number fluctuations. 

As one can see, problem (\ref{LBS2p1}) - (\ref{LBS2p2B4}) and hence total flux $Q_T$ through a particle or a chain of particles, (\ref{QT1}) or (\ref{QT2}), are invariant under the transformation of the particle dimension $R$ provided that the angular size of the bridge $\theta_0$ is fixed. In what follows, we change to non-dimensional description by normalising length scales by the average radius $R_0$ of the particles in the sample and pressure by the characteristic capillary pressure $p_0=2\gamma\cos\phi_c/R_0$. The flux $Q_T$ will be normalised by the characteristic value  
$$ 
Q_0=p_0 \delta_R \frac{\kappa_m}{\mu}\frac{\bar{U}_2-\bar{U}_1}{\bar{L}^B_z}
$$ inspired by the analytical result (\ref{QT1}) and by the non-dimensional sample box surface area $S_0=\bar{L}_x^B \bar{L}_y^B$, where non-dimensional quantities $\bar{L}_{x,y,z}^B=L_{x,y,z}^B/R_0$ and $\bar{U}_{1,2}=U_{1,2}/p_0$. The latter normalisation allows to bring simulation results in slightly different geometric settings, as is detailed in Table \ref{Table1}, into equivalent conditions suitable for comparison, that is basically providing the non-dimensional permeability $\bar{K}=\frac{K}{\kappa_m}\frac{R_0}{\delta_R}$.

\begin{table}[h]
\centering
\resizebox{0.69\textwidth}{!}{
\begin{tabular}{| c | c | c | c | c | c | c | c | c |}
\hline
 &  \multicolumn{8}{| c |}{ Parameters of the configurations} \\ \cline{1-9}
 Configuration & $N_1$  & $N_2$	&	$N_3$ & $\bar{R}/R_0$ & $\phi_S$ (\%) & $L_x^B/R_0$ & $L_y^B/R_0$	& $L_z^B/R_0$		\\ \hline
1 & 5 &  6  & 4 & 1 & 52 & 5.2 & 4.6 & 5.5\\ \hline
2 & 5 &  7  & 5	& 1 & 47 & 5.7 & 4.8 & 4.9\\ \hline
3 & 5 &  4  & 4	& 1 & 47 & 5.2 & 3.7 & 5.3\\ \hline
\end{tabular} }
\caption{Tabulated values of the system parameters in the random configurations used in the study. Here, $N_1, N_2$ and $N_3$ are the number of particles in the configuration with radii $R_1=1.3\, R_0, R_2=R_0$ and $R_3=0.7\, R_0$ respectively, $\phi_S$ is porosity of the sample and $\bar{R}=\frac{N_k R_k}{\sum_{j=1}^3 N_j}$.  
\label{Table1}}
\end{table}

\begin{figure}[ht!]
\begin{center}
\includegraphics[trim=-1.5cm 0.2cm 1cm 0.5cm, width=0.9\columnwidth]{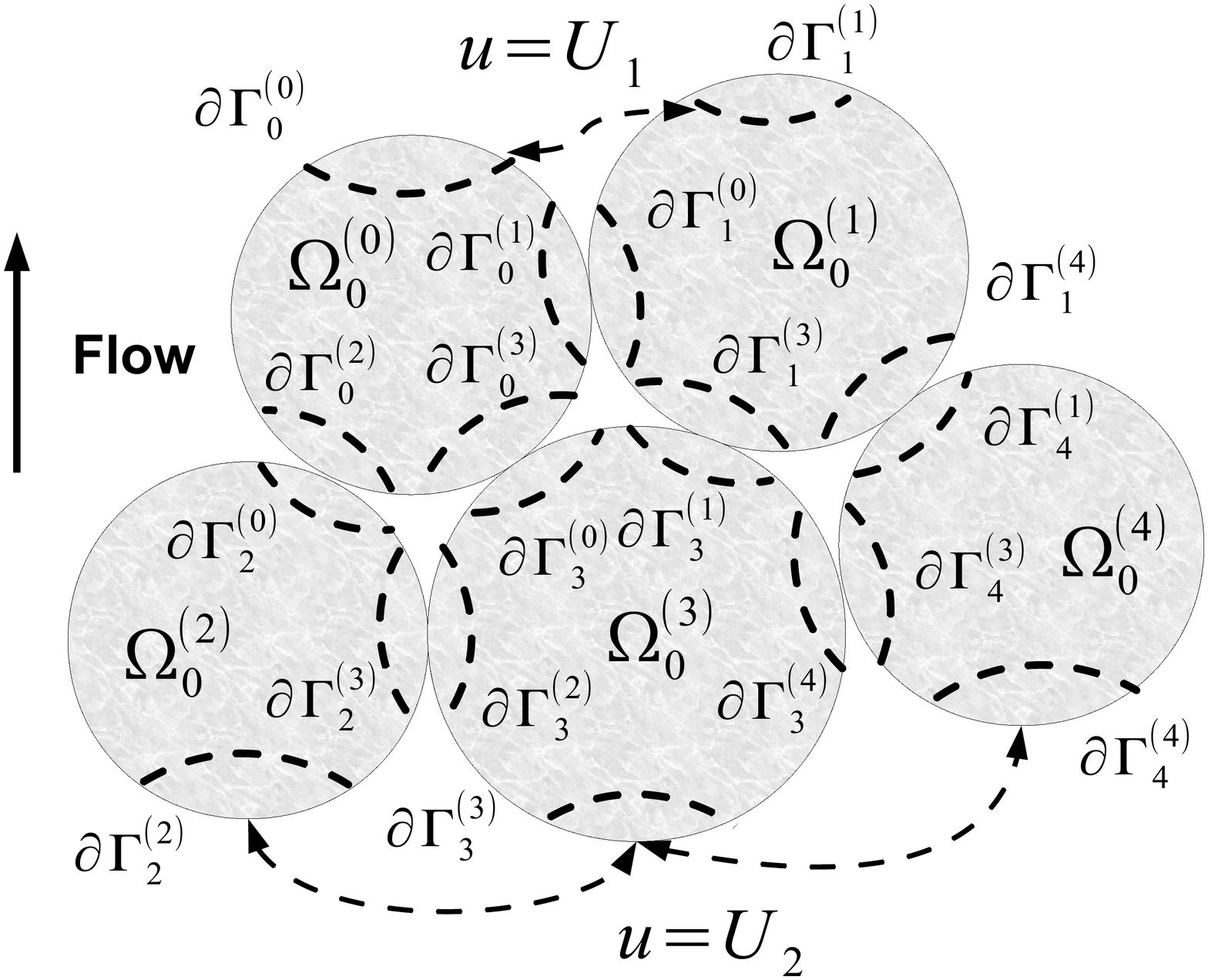}
\end{center}
\caption{Illustration of the particle sample and the flow domains.} 
\label{Fig6}
\end{figure}

\begin{figure}[ht!]
\begin{center}
\includegraphics[trim=-0.5cm 0.2cm 1cm 0.5cm, width=0.6\columnwidth]{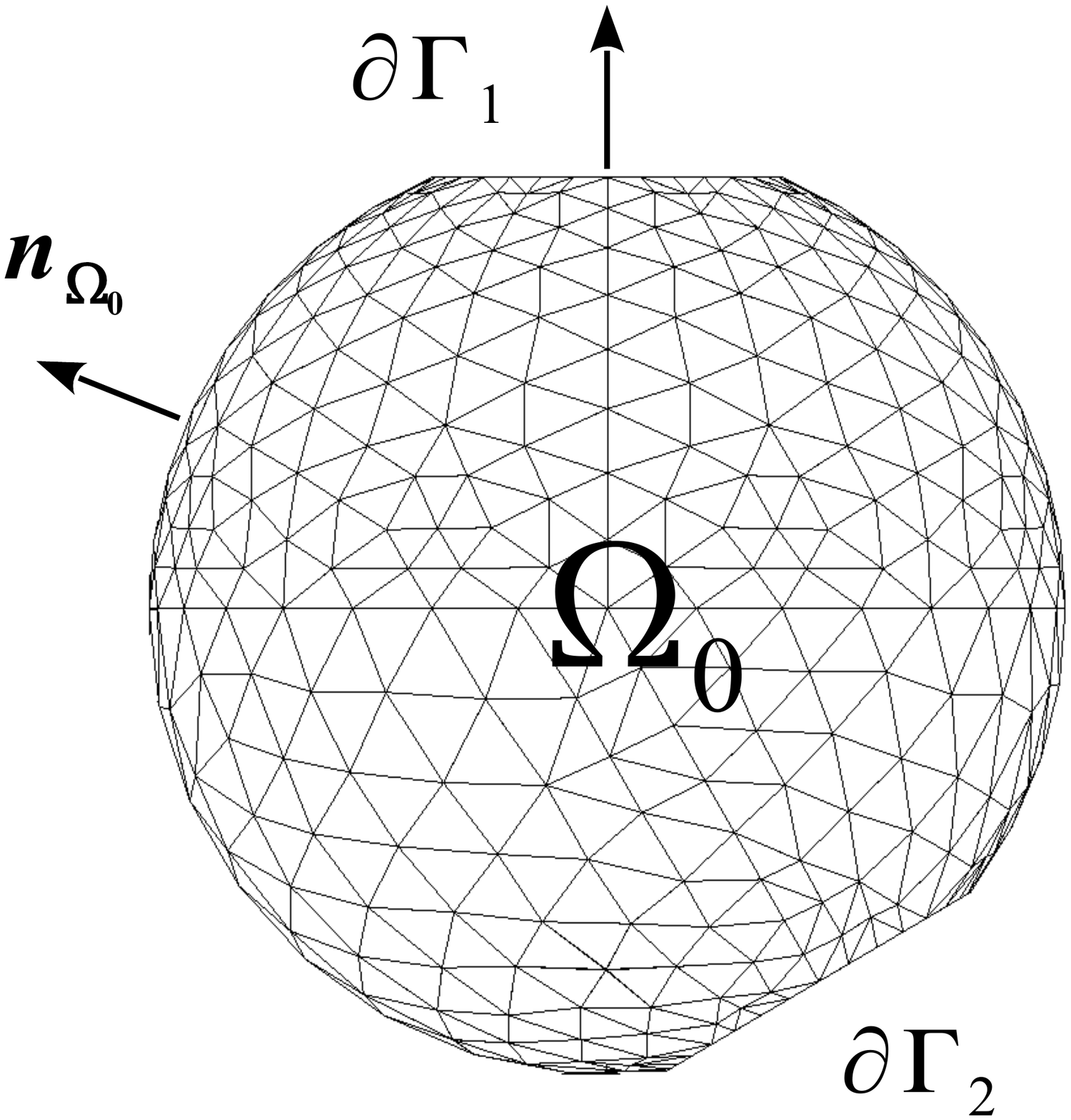}
\end{center}
\caption{Illustration of the tessellated flow domain of a particle for the surface finite element method.} 
\label{Fig5}
\end{figure}

\begin{figure}[ht!]
\begin{center}
\includegraphics[trim=0.5cm 0.2cm 1cm -0.5cm, width=0.9\columnwidth]{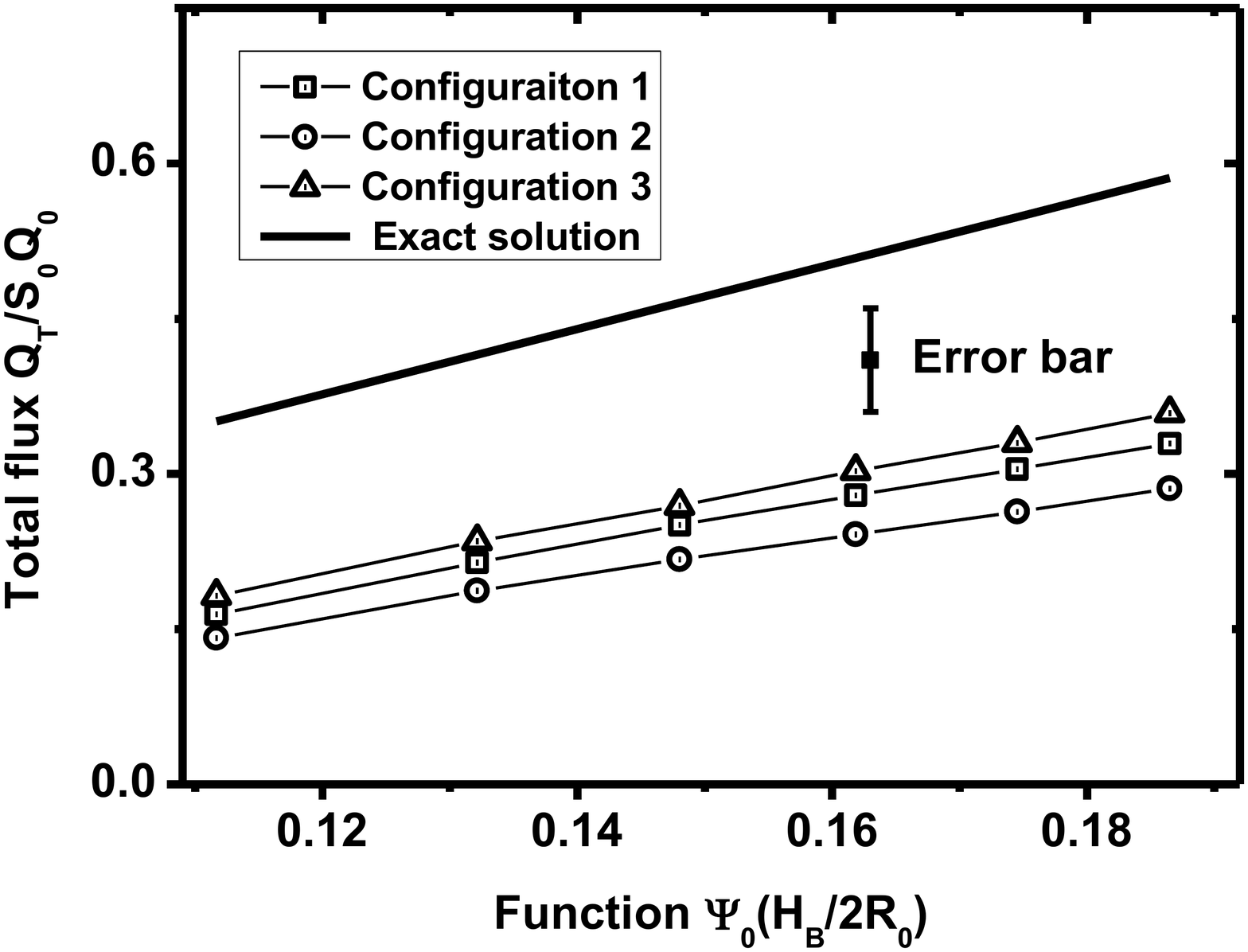}
\end{center}
\caption{Reduced total flux $Q_T/S_0 Q_0$, $\displaystyle Q_0=p_0 \delta_R \frac{\kappa_m}{\mu}\frac{\bar{U}_2-\bar{U}_1}{\bar{L}^B_z}$ and $S_0=\bar{L}_x^B \bar{L}_y^B$, as a function of $\displaystyle \Psi_0(H_B/2R_0)$. The error bar indicates the statistical error, which is expected due to the fluctuations of the number of particles in the samples.}  
\label{Fig8}
\end{figure}

Schematically, the simulation domains for the Laplace-Beltrami problem are shown in Fig. \ref{Fig6}. As in the previous case of particles coupled in a chain, there are internal, common boundaries, contours $\partial \Gamma_{k}^{(l)}$, $k\neq l$, where the continuity boundary conditions are applied and external boundaries, contours $\partial \Gamma_k^{(k)}$, where the Dirichlet boundary conditions are set to generate a flow through the system. The pressure value on the contours facing the bottom of the simulation box (for example, $\partial\Gamma_2^{(2)}, \partial \Gamma_3^{(3)}$ and $\partial \Gamma_4^{(4)}$ in Fig. \ref{Fig6}) is set to $U_2$ and on the contours facing the top side of the pack (for example, $\partial\Gamma_0^{(0)}$ and $\partial \Gamma_1^{(1)}$ in Fig. \ref{Fig6}) is set to $U_1$. The values of the boundary pressure $U_{1,2}$ were identical in the simulations involving different configurations.  

Geometrically, the external boundary contours are oriented in the flow direction, as is illustrated in Fig. \ref{Fig6}. While this particular orientation seems to be arbitrary or may even look artificial, within the statistical approach, the choice of the boundary contour orientation should not render any excessive (in excess of the statistical errors due to the particle number fluctuations) influence upon the results, that is the value of the total flux and the 'macroscopic' permeability. A posteriori, one can see that this seemed to be the case, Fig. \ref{Fig8}, as in different configurations, Table \ref{Table1}, the resultant curves are close and parallel to each other. 

 There are two main questions, we would like to answer in this part of the study. First, how does permeability of the particle sample depend on the composition? Basically, how strong are there fluctuations? Secondly, what is the contribution of the tortuosity effects? To obtain statistically meaningful results, we consider several randomly generated configurations, as is summarised in Table \ref{Table1}. We would like to stress here, that all configurations have been cut off from statistically independent particle distributions generated with the help of random initial distributions of larger number of particles, as we have described. 

As before, we are going to find a weak solution to a system of the Laplace-Beltrami equations
$$ 
\Delta_{\Omega^{(k)}_0} u_k = 0 
$$
defined on each particle domain $\Omega^{(k)}_0$, as in in Fig. \ref{Fig6}. On the internal boundaries of the domains we set up continuity conditions, for example on $\partial \Gamma_3^{(2)}$ and $\partial \Gamma_2^{(3)}$ 
\begin{equation}
\label{Cont1-G}
u_2\left. \right|_{\partial \Gamma_2^{(3)}}=u_3\left. \right|_{\partial \Gamma_3^{(2)}}=const,
\end{equation}

\begin{equation}
\label{Cont2-G}
\oint_{\partial\Gamma_2^{(3)}} \nabla u_2 \cdot  {\bf n}_{s_2}|_{\partial\Gamma_{2}^{(3)}}\, dl =-\oint_{\partial\Gamma_3^{(2)} }\nabla u_3 \cdot  {\bf n}_{s_3}|_{\partial\Gamma_{3}^{(2)}}\, dl.
\end{equation}
While on a few external boundaries, Dirichlet boundary conditions are set.

The numerical solution allows to calculate the total flux through the system by summing up the fluxes passing through the external contours, where the Dirichlet boundary conditions are set, either at the top of the pack or at the bottom using (\ref{TotalInt}). The results are summarised in Fig.  \ref{Fig8}. 

Remarkably, the reduced flux $Q_T/ S_0 Q_0$ as a function of 
$$ 
\Psi_0(H_B/2R_0)=\ln^{-1} \left(\frac{1+ \sqrt{1-\left(\frac{H_B}{2R_0}\right)^2} }{1- \sqrt{1-\left(\frac{H_B}{2R_0}\right)^2}} \right),
$$
\vspace*{0.2cm}
\noindent where $H_B$ is the bridge size $H_B=2R_0\sin\theta_0$, behaves linearly in all configurations. This behaviour mirrors the flux dependence observed in azimuthally symmetric analytical solutions, see (\ref{QT1}) or (\ref{QT2}). The variations in the dependencies between different configurations are observed to be well within the statistical error expected in this case, error bar in Fig \ref{Fig8}. At the same time, a comparison with a similar, but a regular arrangement, as in Fig. \ref{Fig3} at $R=R_0$ demonstrates that there is a clear cut contribution from the effects of tortuosity, solid line in Fig. \ref{Fig8}. 

Indeed, given identical porosity ($\phi\approx 50\%$) and mean particle size ($R/R_0=1$) in the regular, symmetric and randomly generated configurations, the normalised flux values differ by a factor of two, which is consistent with the tortuosity values obtained in porous media in different conditions and configurations~\cite{Tortuosity-Review2013}. For example, both hydraulic $\tau_h$ and diffusive $\tau_d$ tortuosity estimated in unsaturated porous media using different permeability models (often used in applications, for example~\cite{Mualem1976, Mualem1978}) was found in between $1.5 \le \tau_{h,d}\le 2$ at $\phi=50\%$~\cite{Tortuosity-Review2013}.

It is important that the result, that is the ratio of the total flux in the random and regular configurations does not practically depend on the size of the contour $H_B$, basically the size of the liquid bridge, and hence the value of saturation in the porous media. This implies, that the observed effect is purely down to the distribution of contacts between the particles, but not the particular pathway on each single particle surface. That is, fundamentally, tortuosity in the surface diffusion processes is a geometric factor independent of the particular surface flow regime. At the same time, the pathways, on average, of course, does depend on the bridge size value $H_B$ leading to smaller permeability as the size of the contact area diminishes. This trend is expected, but essentially, the correction to the effective coefficient of diffusion   
$$D(s)\propto \frac{1}{|\ln(s-s_0)|(s-s_0)^{3/2}}$$
is only down to a single universal factor of two representing the tortuosity effects in surface diffusion in particular porous media at low values of saturation. Note, this value is also in agreement with experimental observations and a comparison of the super-fast diffusion model with the data, where the tortuosity effects were estimated to reduce the effective permeability twofold~\cite{Lukyanov2019}.

This is the main result of this study, which can be used in practical applications to calculate permeability in particular porous media. Basically, as the first step, one can calculate permeability of a single, representative element of the media or several elements to obtain some mean value and its dispersion. This way permeability $K(s)$, via (\ref{GPSphere}), and the diffusion coefficient $D(s)$ for the macroscopic model can be established in the first approximation. Macroscopic permeability $K(s)$ or the diffusion coefficient $D(s)$ then should be corrected by the universal factor of two in the macroscopic diffusion model.  

\section{Conclusions}

We have demonstrated that the Laplace-Beltrami method can be used to obtain permeability of particulate porous media at low saturation levels and to estimate contribution from the effects of tortuosity. Essentially, analytical results obtained using azimuthally oriented coupled particles can be used with a universal correcting prefactor to estimate permeability of particle ensembles. That is, from the practical point of view, results obtained by analysing single representative element of particulate porous media can be translated into permeability of a particle composition.  

\bigskip
PS was supported through the Royal Thai Government scholarship. TP was partially supported through the EPSRC grant EP/P000835/1.
\bigskip

\twocolumngrid


\begin{thebibliography}{31}

\bibitem{Bear-Book} \by{Bear, J.} {\it Dynamics of Fluids in Porous Media} (Dover, 1972)

\bibitem{Herminghaus-2005} \by{Herminghaus, S.} \paper{Dynamics of wet granular matter} \jour{Adv. Phys.} \vol{54} \pages{221} \yr{2005}

\bibitem{Herminghaus-2008} \by{Scheel, M.; Seemann, R.; Brinkmann, M.; Michiel, M.D.I.; Sheppard, A.; Breidenbach, B. and Herminghaus, S.} \paper{Morphological clues to wet granular pile stability} \jour{Nature Mater.} \vol{7} \pages{189} \yr{2008}

\bibitem{Herminghaus-2008-2} \by{Scheel, M.; Seemann, R.; Brinkmann, M.; Michiel, M.D.I.; Sheppard, A. and Herminghaus, S.} \paper{Liquid distribution and cohesion in wet granular assemblies beyond the capillary bridge regime} \jour{J. Phys. Condens. Matter} \vol{20} \pages{494236} \yr{2008}

\bibitem{Lukyanov2012} \by{Lukyanov, A.V.; Sushchikh, M.M.; Baines, M.J. and Theofanous, T.G.}  \paper{Superfast Nonlinear Diffusion: Capillary Transport in Particulate Porous Media} \jour{Phys. Rev. Lett.}  \vol{109} \pages{214501} \yr{2012}

\bibitem{Lukyanov2019} \by{Lukyanov, A.V.; Mitkin, V.V.; Theofanous, T.G. and Baines, M.J.}  \paper{Capillary transport in particulate porous media at low levels of saturation} \jour{J. Appl. Phys.}  \vol{125} \pages{185301} \yr{2019}

\bibitem{Vazquez-Book}  \by{Vazquez, J.L.} {\it The Porous Medium Equation: Mathematical Theory} (Oxford University Press, 2006)

\bibitem{Halsey1998} \by{Halsey, T.C. and Levine, A.J.}  \paper{How Sandcastles Fall} \jour{Phys. Rev. Lett.}  \vol{80} \pages{3141-3144} \yr{1998}

\bibitem{Yost-1998}  \by{Rye, R.R.; Yost, F.G. and O'Toole, E.J.} \paper{Capillary flow in irregular surface grooves} \jour{Langmuir} \vol{14} \pages{3937} \yr{1998}

\bibitem{Tuller-2000} \by{Or, D. and Tuller, M.} \paper{Flow in unsaturated fractured porous media: Hydraulic conductivity of rough surfaces} \jour{Water Resour. Res.} \vol{36} \pages{1165--1177} \yr{2000}

\bibitem{Whitaker-1969}  \by{Whitaker, S.} \paper{Advances in Theory of Fluid Motion in Porous Media} \jour{Ind. Eng. Chem.} \vol{61} \pages{14--28} \yr{1969}

\bibitem{Penpark2018}  \by{Sirimark, P.; Lukyanov, A.V. and Pryer, T.} \paper{Surface permeability of porous media particles
and capillary transport} \jour{Eur. Phys. J. E} \vol{41} \pages{106} \yr{2018}

\bibitem{Alshibli2004} \by{Alshibli, K.A. and Alsaleh, M.I.}  \paper{Characterizing Surface Roughness and Shape of Sands Using Digital Microscopy} \jour{J. Comput. Civil Eng.}  \vol{18} \pages{36-45} \yr{2004}

\bibitem{Nanoscale2016}  \by{Chen, J.; Wang, C.;  Wei, N.;  Wan, R.  and  Gao, Y.} \paper{3D flexible water channel: stretchability of nanoscale water bridge} \jour{Nanoscale} \vol{8} \pages{5676--5681} \yr{2016}

\bibitem{Tortuosity1937}  \by{Carman, P.C.} \paper{Fluid flow through granular beds} \jour{Trans. Inst. Chem. Eng.} \vol{15} \pages{150--166} \yr{1937}

\bibitem{Tortuosity1961}  \by{Lorenz, P.B.} \paper{Tortuosity in Porous Media} \jour{Nature} \vol{189} \pages{386--387} \yr{1961}

\bibitem{Tortuosity-Review2013}  \by{Ghanbarian, B.; Hunt, A.G.; Ewing, R.P. and Sahimi, M.} \paper{Tortuosity in Porous Media: A Critical Review} \jour{Soil Sci. Soc. Am. J.} \vol{77} \pages{1461--1477} \yr{2013}

\bibitem{Willett2000} \by{Willett, C.D.; Adams, M.J.; Johnson, S.A. and Seville, J.P.K.} \paper{Capillary Bridges between Two Spherical Bodies} \jour{Langmuir} \vol{16} \pages{9396--9405} \yr{2000}

\bibitem{Dziuk1988}  \by{Dziuk, G.} \paper{Finite elements for the Beltrami operator on arbitrary surfaces} \jour{Partial Differential Equations and Calculus of Variations} \vol{1357} \pages{142--155} \yr{1988}

\bibitem{AMS2005}  \by{Pigola, S. and Rigoli, M. and Setti, A.G.} \paper{Maximum principles on Riemannian manifolds and applications} \jour{Memoirs of the American Mathematical Society} \vol{174} \pages{1--36} \yr{2005}

\bibitem{Dziuk2013}  \by{Dziuk, G. and Elliott, C.M.} \paper{Finite element methods for surface PDEs} \jour{Acta Numerica} \vol{22} \pages{289--396} \yr{2013}

\bibitem{Mualem1976}  \by{Mualem, Y.} \paper{A new model for predicting the hydraulic conductivity of unsaturated porous media} \jour{Water Resour. Res.} \vol{12} \pages{513--522} \yr{1976}

\bibitem{Mualem1978}  \by{Mualem, Y.} \paper{Hydraulic conductivity of unsaturated porous media: Generalized macroscopic approach} \jour{Water Resour. Res.} \vol{14} \pages{325--334} \yr{1978}

\end{thebibliography}
\end{document}